\documentclass[usenatbib]{mnras}
\usepackage[T1]{fontenc}
\usepackage{newtxtext,newtxmath}

\usepackage{float}
\usepackage{graphicx}
\usepackage{booktabs}
\usepackage{amsmath}
\usepackage{multirow}
\usepackage{hyperref}

\newcommand{\jwst}{\textit{JWST}}

\title[PAH Emission in the Pillars of Creation]{Spatial analysis of PAH molecules in the Pillars of Creation using \jwst{}}

\author[Pranav R. Iyengar]{
Pranav R. Iyengar$^{1}$\thanks{E-mail: pri24@imperial.ac.uk}\\
$^{1}$Department of Physics, Imperial College London, London SW7 2AZ, UK
}

\date{Accepted XXX. Received YYY; in original form ZZZ}
\pubyear{2025}

\begin{document}
\label{firstpage}
\maketitle
\begin{abstract}
I present a spatially resolved analysis of the polycyclic aromatic hydrocarbon (PAH) population in the Pillars of Creation within the Eagle Nebula (M16) using James Webb Space Telescope (JWST) MIRI and NIRCam imaging. By using mid infrared PAH sensitive bands, I derive resolved maps of PAH size and ionization state across the pillars and connect these directly to variations in the radiation field and gas structure. I present the first spatial maps of PAH ionization and size in the Pillars of Creation. The analysis reveals clear internal gradients that show the PAH population is strongly reshaped by local conditions within the cloud, such as the local radiation intensity and orientation of the nebular structure. The intensely radiated regions show a neutral and large PAH population, possibly due to electron recombination in these regions. I measure a mean PAH size of $198 \pm 1.22$ for M16 and use the resolved emission structure to obtain a first order estimate of the electron density in the molecular cloud. These results provide direct evidence that PAH properties in M16 are governed by the interplay between radiation and density on sub-cloud scales, demonstrating the power of JWST imaging to probe dust processing in star-forming regions.
\end{abstract}


\section{Introduction}
Polycyclic Aromatic Hydrocarbons (PAHs) are a ubiquitous and crucial component of the interstellar medium (ISM) in galaxies. These large, carbonaceous molecules are the primary carriers of the prominent mid-infrared emission features observed at 3.3, 6.2, 7.7, 8.6, and 11.3 $\mu$m, which can dominate the infrared output of star-forming regions and entire galaxies (\cite{galliano_variations_2008}). Through the photoelectric effect on their surfaces, PAHs are also the dominant heating agent for neutral atomic gas in the ISM, playing a critical role in regulating the thermal balance of Photodissociation Regions (PDRs) (\cite{micelotta_polycyclic_2010}). Understanding the life cycle of these molecules, their formation, evolution, and destruction, is therefore fundamental to our understanding of the physics and chemistry of the ISM and the processes that govern star formation.

The physical properties of the PAH population are not static; they are actively sculpted by the local environment. The intense far-ultraviolet (FUV) radiation fields produced by young massive stars are the main driver of this evolution (\cite{allain_photodestruction_1996}; \cite{draine_infrared_2007}; \cite{tielens_interstellar_2008}). This stellar feedback initiates a cascade of physical processes, including the ionization of PAHs into cations and the size-dependent destruction of the molecules themselves. Theoretical models and laboratory experiments have established that smaller and more fragile PAH are preferentially destroyed by energetic photons, a process known as photo-processing, which should lead to an increase in the average size of the surviving PAH population in irradiated regions (\cite{draine_infrared_2007}; \cite{allain_photodestruction_1996}). However, in the most extreme environments, even large PAHs can be shattered into smaller pieces through photo-fragmentation (\cite{zhen_laboratory_2015}). Although these processes are well-established in theory, spatially resolving their effects within a single, complex astrophysical environment has been a long-standing observational challenge.

The Eagle Nebula (M16), with its iconic "Pillars of Creation," is a well studied star forming region. Mid-infrared spectroscopic studies have confirmed the presence of prominent PAH emission features within the nebula (\cite{2003A&A...409..193U}). The nebula is powered by the young massive stellar cluster NGC 6611, which contains numerous O- and B-type stars that produce a powerful and pervasive FUV radiation field (\cite{guarcello_correlation_2007}). Such intense UV irradiation is known to influence PAH charge state and size distribution in photodissociation regions (PDRs), as demonstrated in experiments by \cite{zhen_laboratory_2015} and spatially resolved studies of regions such as NGC 7023 (\citealt{croiset_mapping_2016}).

The pillars themselves are dense, clumpy structures of molecular gas and dust that are actively sculpted and photo-evaporated by this radiation (\cite{oliveira_star_2006}). This geometry creates sharp, well-defined gradients in the radiation over spatial scales, from the intensely irradiated surfaces of the PDRs to the cold, shielded interiors of the nebula. While spatial variations in PAH properties have been extensively investigated in classical PDRs, detailed mapping of PAH processing within irradiated pillar structures has remained limited, particularly at the angular resolution now achievable with JWST. This makes M16 a particularly well-suited region for resolving the response of the PAH population to strong UV-driven environmental gradients within a single field of view.

Previous observations with telescopes such as the Spitzer Space Telescope were limited by their spatial resolution (\cite{2004ApJS..154....1W}), preventing a detailed analysis of the fine structures within the pillars. The advent of the James Webb Space Telescope (JWST) (\cite{2023PASP..135f8001G}) has revolutionized this field, providing for the first time a necessary combination of high spatial resolution and mid-infrared sensitivity to map the detailed variations in PAH properties on the scales at which they are physically processed.

In this paper, I use JWST's NIRCAM (\cite{2023PASP..135b8001R}) and MIRI (\cite{2020sea..confE.229L}) instruments to to resolve PAH size and ionization diagnostics across the irradiated interfaces of the Pillars of Creation and create the first spatial map of the PAH ionization and size gradients of the Pillars of Creation. By constructing a series of diagnostic ratio maps and correlating them with tracers of the ionized and molecular gas, I isolate spatial transitions in PAH properties across the ionization fronts. I present quantitative radial profiles to trace the evolution of PAHs from the shielded cores to the irradiated surfaces of the pillars and use regional statistics to characterize the distinct physical zones of the nebula. This analysis provides spatially resolved observational constraints on PAH processing within irradiated pillar systems, demonstrating how stellar feedback reshapes the molecular component of the interstellar medium at unprecedented resolutions in one of the galaxy's most iconic star-forming regions. This paper is divided as follows: Section \ref{2} describes the data processing steps. Section \ref{3} describes the methods used to obtain the maps. Section \ref{4} elaborates the results and discussion in section \ref{5}. We finally compare M16 to other nebulae with similar analysis in Section \ref{6}

\section{Data Processing}\label{2}
Observations of JWST (\cite{2023PASP..135f8001G}) using Mid infrared instrument (MIRI) (\cite{2023PASP..135b8001R}) and the Near-InfraRed Camera (NIRCAM) (\cite{2020sea..confE.229L}) were obtained from MAST with programme ID 2739 (PI: M. Pontoppidan). For MIRI, I used data from filters F770W, F1130W, and F1500W. For NIRCAM, I used F187N, F200W, F335M, F444W, and F470N. All images used were calibrated at level 3. The following subsections describe the various steps taken to prepare these images for scientific use.

\subsection{Stellar removal}
The original Level 3 JWST images contained numerous bright stellar sources that significantly contaminated the diffuse cloud emission. To isolate the PAH signal and enhance the visibility of extended emission from the pillars, it was necessary to remove these stellar components before further analysis.

Stellar sources were identified using the segmentation maps provided with the Level 3 data products. To separate stars from diffuse emission, I filtered objects based on their projected area in pixels. The exact area threshold was determined empirically by visual inspection for each image, since crowding and PSF size varied across filters. In practice, compact sources were effectively removed with this adaptive approach, while extended nebulosity was preserved. A binary mask was generated from these objects and applied to the science image. All masks were inspected manually to ensure that no diffuse pillar emission was inadvertently excised. The masked pixels were subsequently filled using interpolation from neighboring valid pixels to preserve local background structure. As most of the stars were present in the background sky, I interpolated the removed stars to get a complete image. Due to the small number of stars in the actual nebula, interpolation would not affect the size or ionization gradients much. This procedure effectively removed the majority of stars. For brighter or more extended stars exceeding the initial area threshold, the process was repeated with an adjusted segmentation criterion.

\begin{figure*}
    \includegraphics[width=0.45\textwidth]{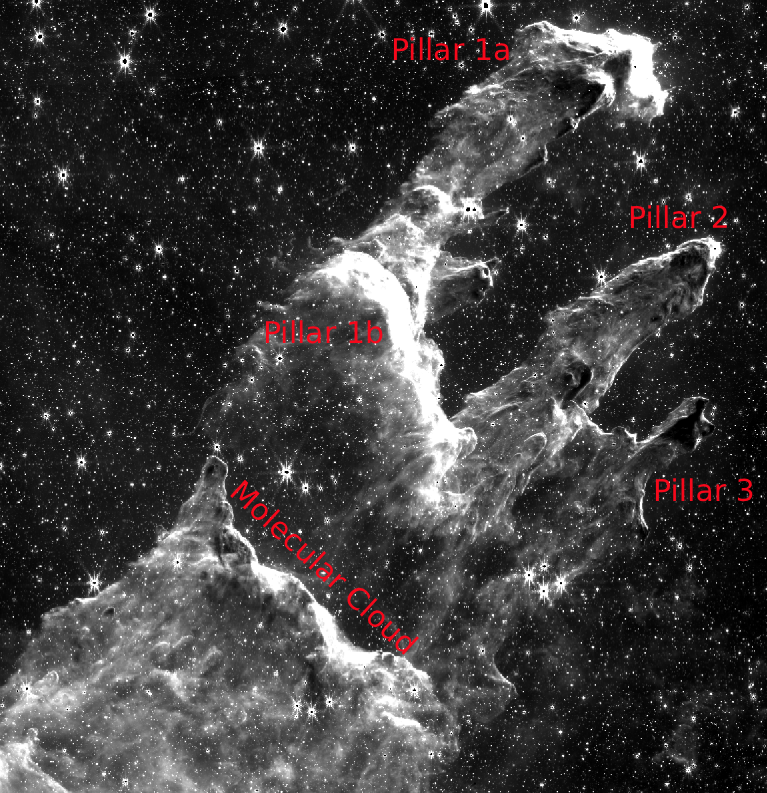}
    \includegraphics[width=0.45\textwidth]{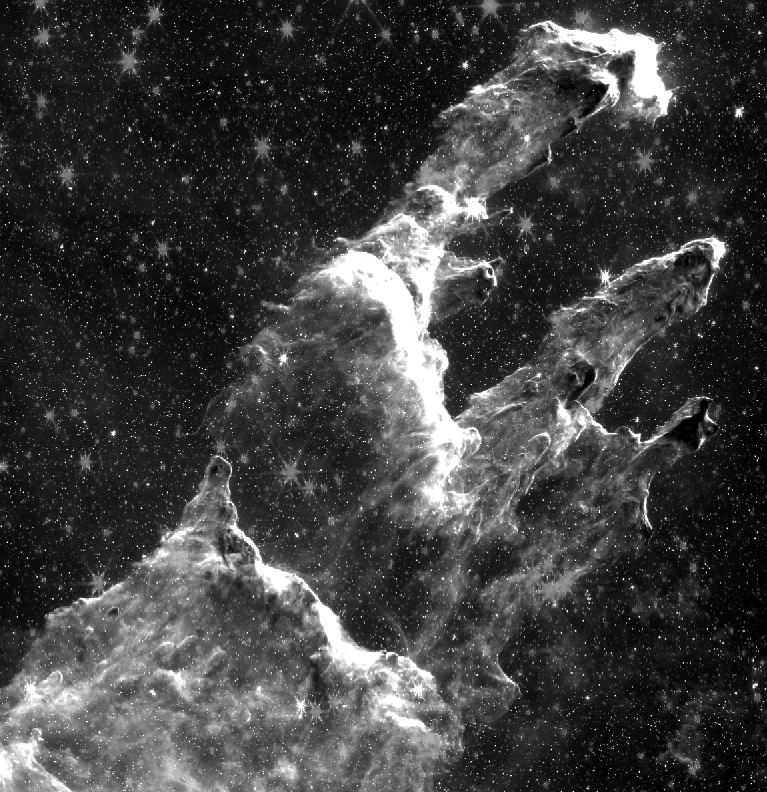}
    \caption{Image depicting the stellar removal process. Left panel image shows the original F335M image with the different regions of the nebula written in red. The right panel shows the image after the star removal process. We can see that most of the bright stars have removed and the gas cloud is better highlighted. }
    \label{fig:f335}
\end{figure*}

In cases where bright stars were present in the science image but absent in the segmentation map, manual masking was performed using region files. These manually defined regions were also interpolated in the same manner. The combined approach successfully mitigated most of the stellar contamination and enabled accurate extraction of diffuse PAH emission features across the pillars. The final star removed images are shown in figure \ref{fig:f335}.

\subsection{PSF Matching}
To ensure accurate pixel-by-pixel comparison between filters with different spatial resolutions, all images were convolved to a common point spread function (PSF). This step is essential for constructing ratio maps and applying diagnostics based on multi-band photometry. Each filter has a distinct native PSF due to variations in wavelength and instrument optics, with longer-wavelength MIRI filters exhibiting broader profiles than their NIRCam counterparts. A PSF matching strategy was adopted in which higher-resolution images were convolved to match the resolution of the broadest filter used in the analysis, which in this case is F1500W. The full widths at half maximum (FWHM) values for every filter is given in Table. The required Gaussian kernel width was calculated from the quadrature difference of the FWHM of the source and target PSFs, i.e.

\begin{table}[ht]
\centering
\caption{FWHM values for each of the filters.}
\label{tab:bkg}
\begin{tabular}{lcc}
\hline
      \textbf{Filter}   & \textbf{FWHM} \\
   & (arcsec)\\
\hline
    F187N & 0.061\\
    F200W & 0.064\\
    F335M & 0.109\\
    F444W & 0.140\\
    F470N & 0.154\\
    F770W & 0.269\\
    F1130W & 0.375\\
    F1500W & 0.488\\
\hline
\end{tabular}
\label{tab:placeholder}
\end{table}

\begin{equation}
    \sigma_{arcsec} = \frac{\sqrt{FWHM_{target}^2-FWHM_{orig}^2}}{2.355}
\end{equation}

Here, the factor 2.355 converts a Gaussian’s FWHM into its standard deviation $\sigma$ since FWHM = $2\sqrt{2ln2}\sigma \approx 2.355\sigma$ (\cite{stetson_daophot_1987}). The resulting $\sigma$ was then converted into pixel units using the image pixel scale. This approach assumes circular Gaussian symmetry, which does not capture the diffraction pattern of JWST PSFs but provides a reasonable approximation for extended diffuse emission. Because this study focuses on large-scale PAH emission morphology rather than precise point-source photometry, the Gaussian approach is sufficient and ensures that artificial gradients are minimized in ratio maps. The image was smoothed by convolving it with a Gaussian kernel using the convolve function from astropy.convolution module in Astropy \citep{2024zndo..13860849A}. Following this procedure, all images were placed on a common resolution scale, minimizing artificial gradients in ratio maps and ensuring the physical reliability of derived quantities.

Background emission was carefully subtracted to isolate the intrinsic PAH signal from the pillars. For each filter image, I selected regions around the nebula free of bright stellar or nebular structures to characterize the local background level. These regions provide an estimate of the diffuse Galactic emission and instrumental residuals that are otherwise superimposed on the target. A median background value was computed and removed from the science images. Table \ref{tab:bkg} shows the background levels subtracted from each of the filters.

\begin{table}[ht]
\centering
\caption{Background levels for each of the filters.}
\label{tab:bkg}
\begin{tabular}{lcc}
\hline
      \textbf{Filter}   & \textbf{Background} \\
       & (MJy/sr)\\
\hline
        F187N & 34.03\\
        F200W & 3.33\\
        F335M & 4.41\\
        F444W & 5.612\\
        F470N & 3.843\\
        F770W & 31.3\\
        F1130W & 85.57\\
        F1500W & 110.4\\
\hline
\end{tabular}
\end{table}

\subsection{Continuum Subtraction}

From the fully reduced and processed NIRCAM and MIRI data, continuum-subtracted
images were created for the 3.3$\mu$m, 7.7$\mu$m and 11.3$\mu$m PAH emission feature. The subtraction method utilizes a shorter and a
longer wavelength filter to derive the continuum in the
emission line filter. 

The continuum in the F335M filter at each pixel in the image was estimated
by interpolating the SED between the F200W
and F444W filters at the location of F335M. The resulting continuum image is then subtracted from the F335M
to derive the 3.3$\mu$m PAH emission line map. 

For the F770W filter, the filters used were F444W and F1500W. These filters were chosen as anchors on the shorter side, F470N is a narrow band filter, and F444W is the nearest available wide band filter on the shorter side. On the longer side, F1130W contains PAH features as well; hence F1500W was used, which has negligible PAH features. 

Similarly, for F1300W, the shorter side used F444W, and the longer side used F1500W. The significant spectral separation between the PAH feature and its chosen continuum anchor can introduce uncertainties or lead to incomplete subtraction; however, this limitation arises from the restricted availability of suitable filters for M16.

In order to remove contamination from hydrogen recombination line emission in the broadband filters, we applied an iterative line-removal technique. The broadband filter F200W contains both continuum emission and potential Pa $\alpha$ contamination at 1.875 microns. This Pa $\alpha$ line is traced by the F187N filter. To isolate the pure dust continuum, I implemented a pixel-wise correction based on the method described by \cite{gregg_feedback_2024} using the narrowband filter F187N, which traces Pa $\alpha$ as a line-only proxy. At each iteration, the estimated line flux was calculated by subtracting the current estimate of the continuum from the narrowband image. This line flux was then scaled by the ratio of filter bandwidths ($\Delta\lambda_{narrow}$/$\Delta\lambda_{broad}$) and subtracted from the original broadband image. The process was repeated until convergence, defined by a fractional change of less than $10^{-4}$ between successive iterations. This iterative approach accounts for residual continuum emission in the narrowband image and ensures accurate removal of line emission from the broadband data. The resulting continuum-isolated image was used in subsequent ratio maps and diagnostics involving warm dust emission. Similar contamination of F444W is also seen by the Br $\alpha$ line at 4.05$\mu$m, but unfortunately, none of our available filters trace this line, hence it has been left.

The continuum level at the PAH band was estimated using a power-law model F $\propto \lambda^{\alpha}$, implemented via linear interpolation in log–log space with base 10. Specifically, the continuum intensity at each pixel was computed as:

\begin{multline}
\log F_{\lambda_\text{target}} = \log F_{\lambda_\text{long}} \\
+ \frac{\log(\lambda_\text{target}) - \log(\lambda_\text{long})}
       {\log(\lambda_\text{short}) - \log(\lambda_\text{long})} 
   \cdot \left[ \log F_{\lambda_\text{short}} - \log F_{\lambda_\text{long}} \right]
\end{multline}

and converted back to linear flux space. A scaling factor (\cite{gregg_feedback_2024}) empirically determined by iteratively comparing the continuum-subtracted results for a range of scale values. The optimal value (Table \ref{tab:cont}) was chosen such that it strikes a balance between
the optimal subtraction of the stars in the field, while
also limiting oversubtraction of the continuum in the
nebular regions. Saturated or bright pixels in the F1130W image were masked before subtraction to prevent contamination. The final continuum-subtracted map reveals the line-only PAH emission feature, with minimal contamination from thermal dust continuum. This method provides a robust approach to isolating PAH features when only limited continuum anchor filters are available.

\begin{table}[ht]
\centering
\caption{Table showing the scaling factor used for continuum subtraction}
\label{tab:cont}
\begin{tabular}{lcc}
\hline
      \textbf{Filter}   & \textbf{Scaling factor} \\
\hline
        F335M & 0.3\\
        F770W & 0.7\\
        F1130W & 0.3\\
\hline
\end{tabular}
\end{table}

\section{Methods}\label{3}

Here, the methods involved in obtaining all the individual maps and analysis diagnostics are discussed. The continuum-subtracted images were used for all the following methods.

\subsection{PAH Ionization State and Size Mapping}
PAH ionization state was mapped using the technique mentioned in \cite{maragkoudakis_probing_2020} (hereon M20). Three PAH features were considered, 3.3$\mu$m, 7.7$\mu$m and 11.0/11.2$\mu$m. Cationic PAH mainly emit in the 11.0$\mu$m band while neutral PAH emit in the 11.2 $\mu$m band (\cite{rosenberg_coupled_2011}). The F1130W band encompasses both these bands. The 3.3$\mu$m is captured by the F335M filter, and 7.7$\mu$m is captured by the F770W filter. 

Before applying the M20 technique, a polygon was drawn around the nebula, including pillars and molecular cloud, and saved as a region file. This region file was used to mask the image, leaving only the nebula and removing the surrounding sky. 

In addition, pixels whose PAH emission is indistinguishable from the sky background were also masked. Because genuine PAH emission is broadband across the mid-IR features but can peak differently among bands depending on charge and size, a two-tier, multi-band criterion was designed to maximize purity without biasing against real emitters with band-dependent strengths. First, a common sky background level was estimated for each PAH image via aperture photometry in the low-surface-brightness region above the molecular cloud and below the pillars (where the flux matches the field background). For masking, a pixel must (i) exceed the lower threshold (the median background level) in all three PAH bands, suppressing false positives from noise fluctuations in any single band; and (ii) exceed an upper threshold of 3× the background in at least one PAH band, ensuring that at least one feature is detected at high significance even if the PAH spectrum is intrinsically stronger in one band than the others. A single per-band threshold would either admit spurious one-band noise spikes (if set low) or systematically reject real PAH emission with asymmetric band strengths (if set high in all bands). The adopted two-tier rule balances completeness and reliability and yields a conservative mask for robust M20 measurements; masked pixels are excluded from all subsequent size statistics.  

M20 also gives different track parameters for different radiation fields. For M16, the O-type stars in NGC 6611 emit strong radiation in the FUV band (FUV) (\cite{anania_novel_2025}). The energy range of FUV lies in the range of 6eV to about 13ev (Considering wavelength between 100nm to 200nm). A mean radiation energy of 10ev was considered for the case of the pillars of creation. The corresponding diagnostic grid for 10ev was used. A pixel is assigned the ionization state of the track it is closest to. 

PAH size mapping was done using the technique mentioned in M20. Two ratios were considered for the analysis, F1130W/F770W and F335M/F770W. Five ionization states were defined, similar to M20: Neutral, Cationic, and a mixture in relative contributions of: (i)75–25 per cent (N75C25), (ii) 50–50 per cent (N50C50), and (iii)25–75 per cent (N25C75). The (11.2+11.0)/7.7–(11.2+11.0)/3.3 plane as mentioned in M20 was used. In this plane, 5 tracks were defined for each of the ionization states. Then a pixel-by-pixel analysis was done, where, depending on the position of the pixel in this plane, it was assigned the ionization of the track it was closest to. Finally, the ionization states were mapped onto the image, with Neutral pixels set to 1, N75C25 to 0.75, N50C50 to 0.5, N25C75 to 0.25, and Cationic to 0.

M20 gives PAH size scaling relations for each of the ionization states. A similar pixel-to-pixel analysis was done. Depending on the ionization state, the appropriate scaling relation was used to estimate the number of carbon atoms in the PAH molecule. Once the PAH sizes were determined, a size cut was applied to remove model edges or noise. For this, the distrbution of the sizes was seen, and a cut was assigned at the point where the tail starts. This value happens to be at 350, and it is at the 99\%ile. A lower cut was also added at 50 to remove the very small sizes.

\subsection{Error analysis}
To quantify the accuracy of the maps an uncertainty analysis of the PAH size.

To quantify per-pixel uncertainties in ionization fraction and PAH size ($N_c$), I performed a Monte Carlo simulation with 1,000 realizations per pixel. Beginning with continuum-subtracted intensity and error maps in F335M, F770W, and F1130W, I generated synthetic realizations by adding Gaussian noise to these images ($\sigma$ from error maps). For each realization, we recalculated the diagnostic ratios and derived PAH properties using our established model. Aggregating over all realizations, we computed the standard deviation of $N_c$, producing corresponding 1$\sigma$ uncertainty maps.

\subsection{Radial Profiles and region-wise analysis}
To investigate the spatial variation of PAH properties across the irradiated surfaces of the pillars, we extracted radial profiles perpendicular to the long axis of each pillar. For this, I defined a series of straight-line cuts across Pillars 1 and 2. Along each cut, pixel values for PAH size was extracted. 

These values were averaged within narrow perpendicular bins to construct smoothed one-dimensional profiles and plotted with corresponding standard deviations. To enable comparison across different cuts and physical scales, the spatial axis of each profile was normalized between 0 and 1. This normalization preserves the qualitative trends while removing dependence on absolute distances. 

For the region-wise analysis, regions around the molecular cloud, Pillar 1a, Pillar 1b, and Pillar 2 were defined to get an idea of PAH size in these regions. In addition, the mean ionization was calculated with the percentage of pixels in each ionization state to get an idea of the distribution of the ionization state in each region.

\section{Results}\label{4}
This section presents the findings derived from the multi-wavelength analysis of the M16 pillars. The initial discussion provides a morphological overview of the principal physical components and their spatial relationships. Subsequently, a quantitative examination of the Polycyclic Aromatic Hydrocarbon (PAH) population is detailed, focusing on its abundance, size, and ionization state. An investigation into the correlations between these properties and the local physical conditions is also presented.

\subsection{Ionizing Radiation and effect of NGC 6611}
\begin{figure}
    \centering
    \includegraphics[width=1\linewidth]{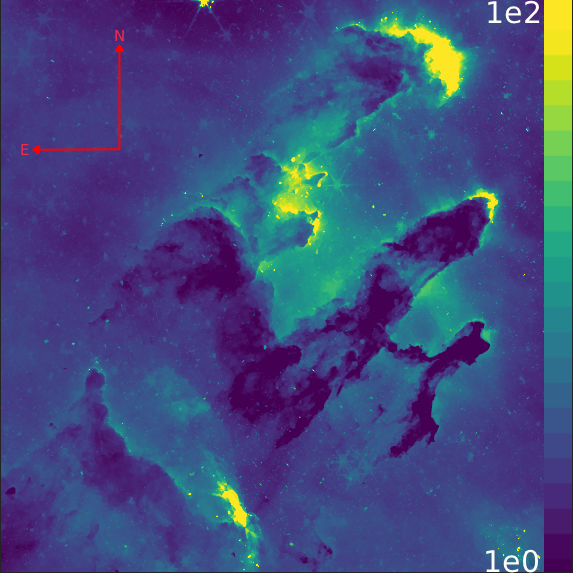}
    \caption{F187N image showing Pa$\alpha$ emission line flux (MJy/sr), which can be treated as a proxy for FUV radiation. The dark blue regions have flux of the order $10^0$ MJy/sr, and the bright yellow regions have flux of the order $10^2$ MJy/sr. The image shows the stellar radiation which is coming from the NGC6611 cluster located above the pillars. North is up and east is to the left.}
    \label{fig:f187}
\end{figure}

The pillars of creation are bombarded with Intense ionizing radiation from the O and B-type stars in the NGC 6611 cluster (\cite{peron_hadronic_2025}). The Paschen-Alpha (Pa$\alpha$) hydrogen recombination line, traced by the F187N filter, can be used as a proxy for the radiation (\cite{peeters_pdrs4all_2023}). This is due to the fact that when in the HII region, free electrons recombine with protons to form hydrogen, electrons cascade down energy levels emitting photons. One such emission is the Pa$\alpha$ emission, whose frequency is in the NIR which can be captured by NIRCAM. Figure \ref{fig:f187} shows the F187N filter data of the Pillars of Creation.

To understand this image and subsequent results its important to focus on the structure of these pillars relative to the cluster. \cite{leod_pillars_2015} suggested a 3D geometrical view of the pillars and the cluster. Using the velocity maps, extinction maps, and the integrated line intensities, they were able to conclude that Pillar 1 is composed of two structures. Pillar 1a (P1a), the top part, which is present behind the cluster with its tail pointed away from us, and Pillar 1b (P1b), the bottom part, which is present in front of the cluster with its tail pointed towards us. Pillar 2 (P2) is a single structure with its tail pointed towards us, while the data from Pillar 3 is inconclusive; they suggest that P3 is placed even closer to us than P2, with its tip pointing towards us. The NGC 6611 cluster is on top of this whole structure. This is pictorially shown in Figure 21 of \cite{leod_pillars_2015}.

This structure can easily be visualized with the NIRCAM F187N image. P1a is behind the cluster, so what we see is the side of the pillar being irradiated. Hence, this region shows high Pa$\alpha$ emission. P1b is situated relatively far away from the stars compared to the other pillars, hence it is not that intensely irradiated as the other pillars. P2 is in front of the cluster. In the F187 image, we can see that the emission seems to be coming from the region behind with respect to our line of sight. This observation supports the 3D structure proposed by \cite{leod_pillars_2015} as the region facing the cluster is what is giving of the Pa$\alpha$ emission. P3 gives off even lesser emission as it is placed much further away with its tip facing us.

\subsection{Photodissociation Regions and RGB Composite Analysis}

A photodissociation region (PDR) is the boundary layer on the surface of a cold, dense molecular cloud where it is irradiated by far-ultraviolet (FUV) photons from nearby massive stars (\cite{salama_pahs_2008}). It marks the transitional interface between the ionized H II region and the neutral molecular gas. In such regions, FUV photons dissociate molecules and heat the gas without fully ionizing hydrogen.

\begin{figure}
    \centering
    \includegraphics[width=1\linewidth]{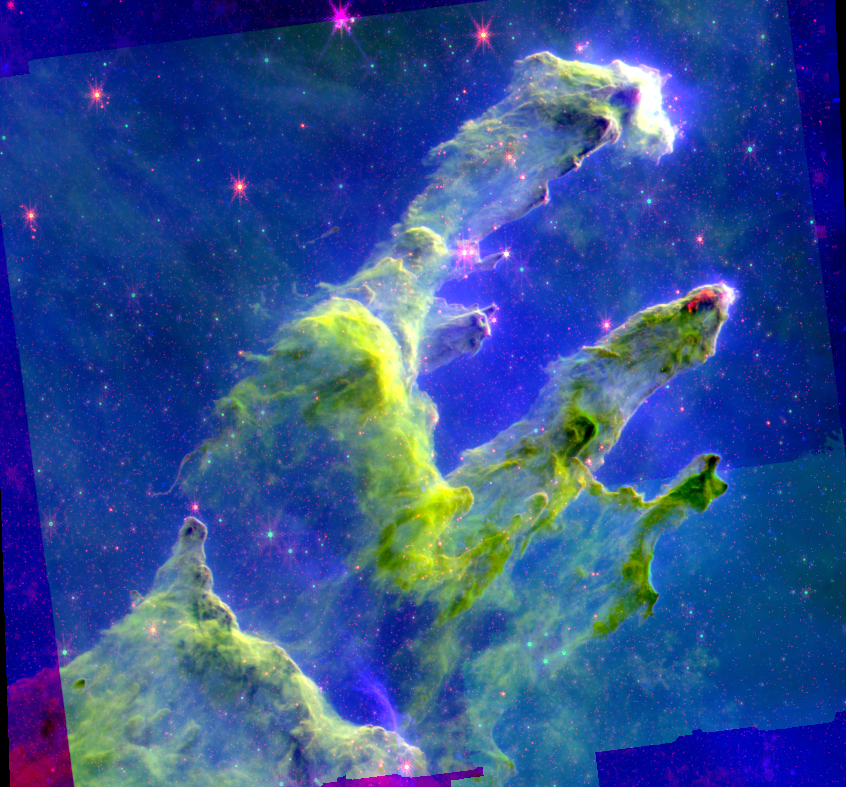}
    \caption{RGB composite of the Pillars of Creation in the Eagle Nebula (M16) observed with JWST. Blue corresponds to Filter NIRCAM F187N which shows the FUV radiation, red to NIRCAM F470N tracing the molecular gas, and green to the MIRI F770W which traces the 7.7$\mu$m PAH feature. The image highlights spatial variations in PAH emission along the irradiated surfaces of the pillars, with enhanced PAH intensity at the ionization fronts facing the central OB cluster. North is up and east is to the left.}
    \label{fig:rgb}
\end{figure}

In the case of M16, I identify three primary physical components using JWST wavebands: ionized gas, molecular gas, and PAH emission. To disentangle these components and visualize their spatial interplay, a three-color composite image was constructed (Figure~\ref{fig:rgb}), using:
\begin{itemize}
  \item F187N (Pa\,$\alpha$; ionized gas) in blue,
  \item F470N (H$_2$ emission; warm molecular gas) in red,
  \item F770W (7.7 $\mu$m PAH feature) in green.
\end{itemize}

This composite image reveals key structural characteristics of the region. The area all around the nebula are blue, indicative of strong Pa\,$\alpha$ emission from ionized hydrogen, the HII region. 

The pillars themselves exhibit classic PDR characteristics. Regions appearing yellow (red + green) correspond to warm molecular gas mixed with PAH emission interacting with the radaition, a signature of PDR. At the very tips of the pillars, white regions emerge which are again PDRs, where all three components—ionized gas, molecular gas, and PAHs—coexist. This confluence is especially apparent at the tips of Pillar 1a and Pillar 2, indicating intense UV bombardment and active photochemical processing.

Interestingly, Pillar 2 appears more red-dominated compared to Pillar 1. This may be due to its orientation: the side facing the observer is likely shielded from the cluster, reducing the intensity of ionizing radiation and resulting in dominant H$_2$ emission with weaker Pa\,$\alpha$ and PAH features. Also it might be possible that the gas in Pillar 2 is modenserhan in Piliar 1a.

Below the pillars, the top surface of the molecular cloud appears yellow, marking a well-defined PDR. Beneath this, redder tones dominate, indicating shielded molecular regions. Further below, areas that appear predominantly green likely represent zones where PAHs are excited but the molecular gas is either too cold to emit or too diffuse, possibly due to double shielding by both the pillars and the molecular surface PDR.

This RGB composite provides a compelling visualization of how stellar feedback from NGC 6611 shapes the morphology and physical conditions within the Eagle Nebula.

\subsection{PAH Ionization}

\begin{figure}
    \centering
    \includegraphics[width=1\linewidth]{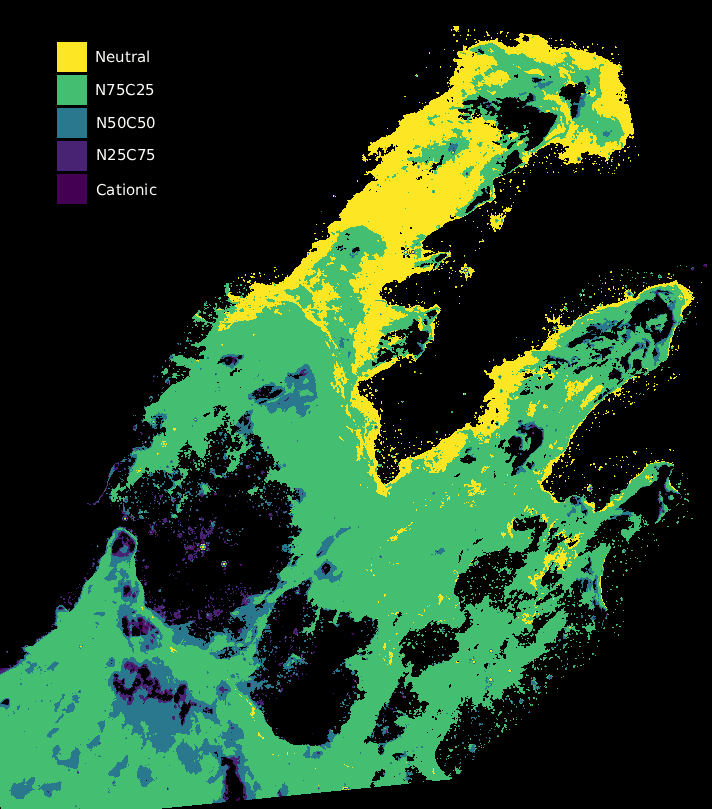}
    \caption{Spatial Map showing the distribution of different ionization states in the nebula. Yellow corresponds to neutral states which populate Pillar 1, and skins of Pillar 2. While blue regions represent cationic ionization states, which are located deeper in the nebula, far away from the intense steallr radiation.}
    \label{fig:ion}
\end{figure}

The ionization map is presented in Figure \ref{fig:ion}. This is plotted in the regions showing PAH emission; the regions with negligible PAH emission are masked. Pillar 3 is not considered in further analysis as there seems to be a rectangular artifact that is seen in the RGB image and it subsequently affects all further maps.

Two major things can be noticed. First, most of the nebula has an ionization state of N75C25 (can be seen in the legend of Figure \ref{fig:sizedis}. Secondly, the pillars, especially Pillar 1a, have a majorly pure neutral ionization state, while the molecular cloud is more cationic in nature. This indicates that highly radiated regions are more neutral than the shielded regions. This may seem counterintuitive at first glance, but it can be explained; which will be discussed in the Discussion section.

\subsection{PAH Molecular Sizes}

\begin{figure}
    \centering
    \includegraphics[width=1\linewidth]{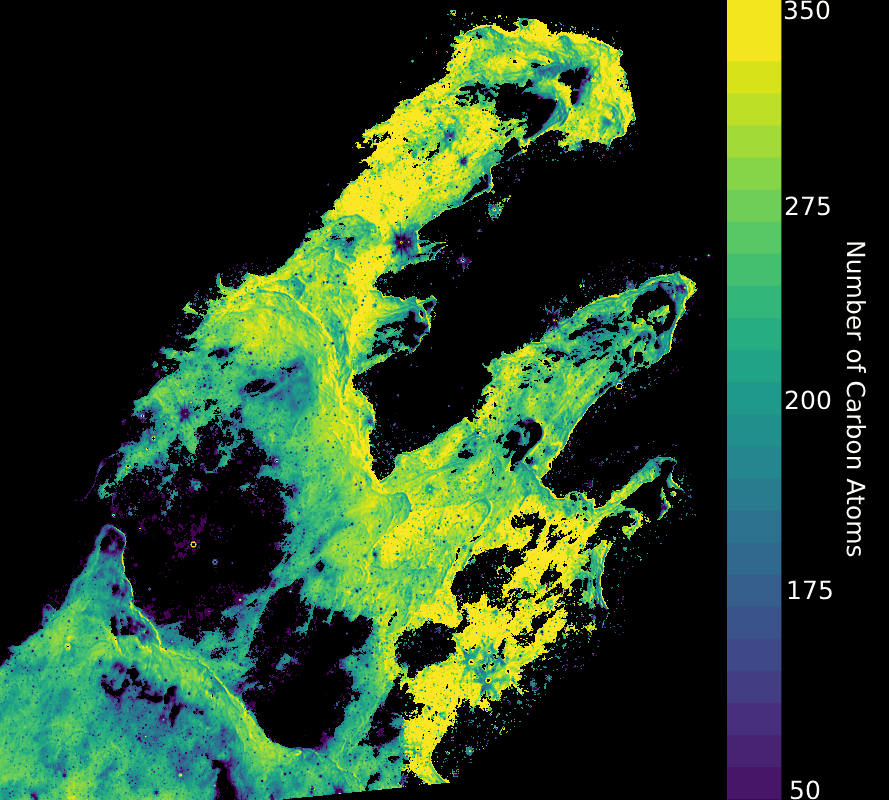}
    \caption{Spatial map showing the size of PAH molecules in different regions of the nebula. The number of carbon atoms represent the size of the PAH molecule. The map shows distinct gradients in the size, with Pillar 1 showing molecules with $N_c$ (Number of carbon atoms) > 300, while the molecular cloud having molecules with $N_c \sim 200$.}
    \label{fig:size}
\end{figure}

Using the technique described in M20 on the ionization map, the molecular sizes of PAH are derived in terms of the number of carbon atoms. The PAH size map is shown in Figure \ref{fig:size}.

The most prominent feature is the presence of the largest PAHs in regions of high radiation, such as Pillar 1a. These regions correspond to the Photodissociation Regions (PDRs) where the molecular cloud is directly interacting with the stellar radiation field.

Pillar 1a has larger PAHs than Pillar 2, as the radiation hitting the side of Pillar 1a facing us is more than the side of Pillar 2 facing us. 

The smallest PAHs are present in the parts of the nebula farther away from the star cluster. As one goes deeper into the nebula away from the stars, the size of PAH reduces. Pillar 2 has a smaller PAH compared to Pillar 1a and Pillar 1b, as it is farther away from the stars. The molecular cloud is both far away and shielded by the pillars, leading to much smaller PAHs. Pillar 3 seems to have uniformly distributed large PAHs; this is due to an artifact mentioned earlier.

A 1000 iteration Monte Carlo simulation was run to get the uncertainty values in the PAH size. The Mean size of PAH in the Pillars of Creation is 198$\pm$1.22 Carbon atoms with a standard deviation of 78.

\begin{figure}
    \centering
    \includegraphics[width=1\linewidth]{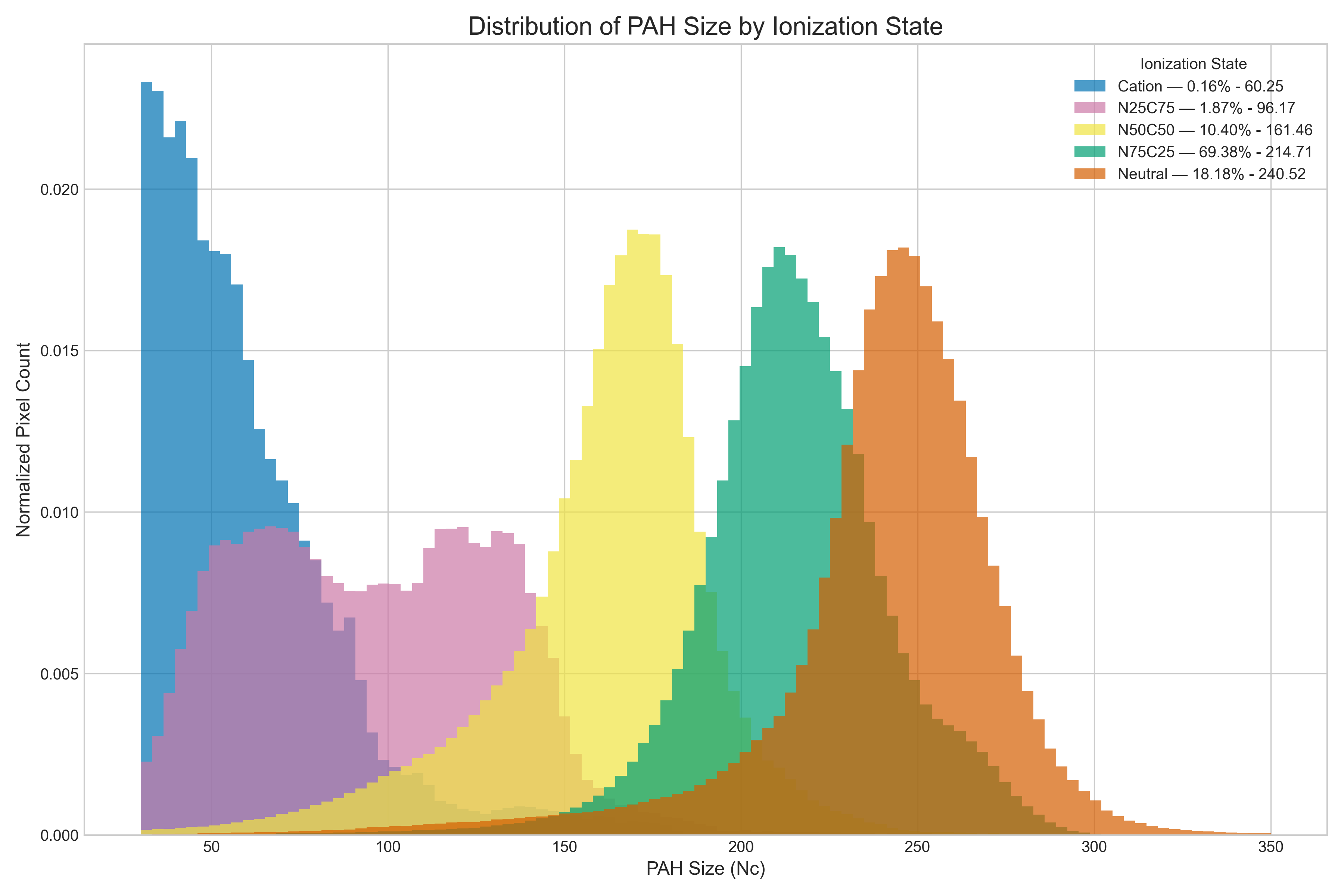}
    \caption{Distribution of ionization state with respect to size. The legend shows the percentage of pixels in each ionization state and the mean PAH size for each ionization state. Cationic state(Blue) having 0.16$\%$ of pixels with mean size of 60.25 carbon atoms, N25C75 (pink) having 1.87$\%$ of pixels with mean size of 96.17 carbon atoms, N50C50 (Yellow) having 10.40$\%$ of pixels with a mean size of 161.46 carbon atoms, N75C25 (Green) having 69.38$\%$ of pixels with mean size of 214.71 carbon atoms, Neutral state (Orange) having 18.18$\%$ of pixels with mean size of 240.52 carbon atoms.}
    \label{fig:sizedis}
\end{figure}

A normalized histogram of PAH sizes for each nominal ionization state is presented in Figure \ref{fig:sizedis}. A distinct trend emerges: the population of pixels classified as Cationic ($N_c$=0.0), found in the most shielded regions, exhibits a distribution peaked at the smallest size ratios. Conversely, the population classified as Neutral ($N_c$=1.0) is larger in size on average and more abundant. As the fraction of neutral molecules increases, the mean size increases. All the distribution shows a Gaussian type of distribution except the N25C75 state, which has a more spread out distribution with two peaks indicating the existence of two populations of the same ionization state. Using a GMM model, these two populations were separated into two Gaussian distributions, and then their spatial distribution was visualized. The population with the larger mean size mainly exists inside the molecular cloud, while the population with the smaller mean size mainly exists as scattered pixels in the regions masked out with negligible PAH emission.

\subsection{Quantitative Profiles and Regional Variations}

\begin{figure}
    \centering
    \includegraphics[width=1\linewidth]{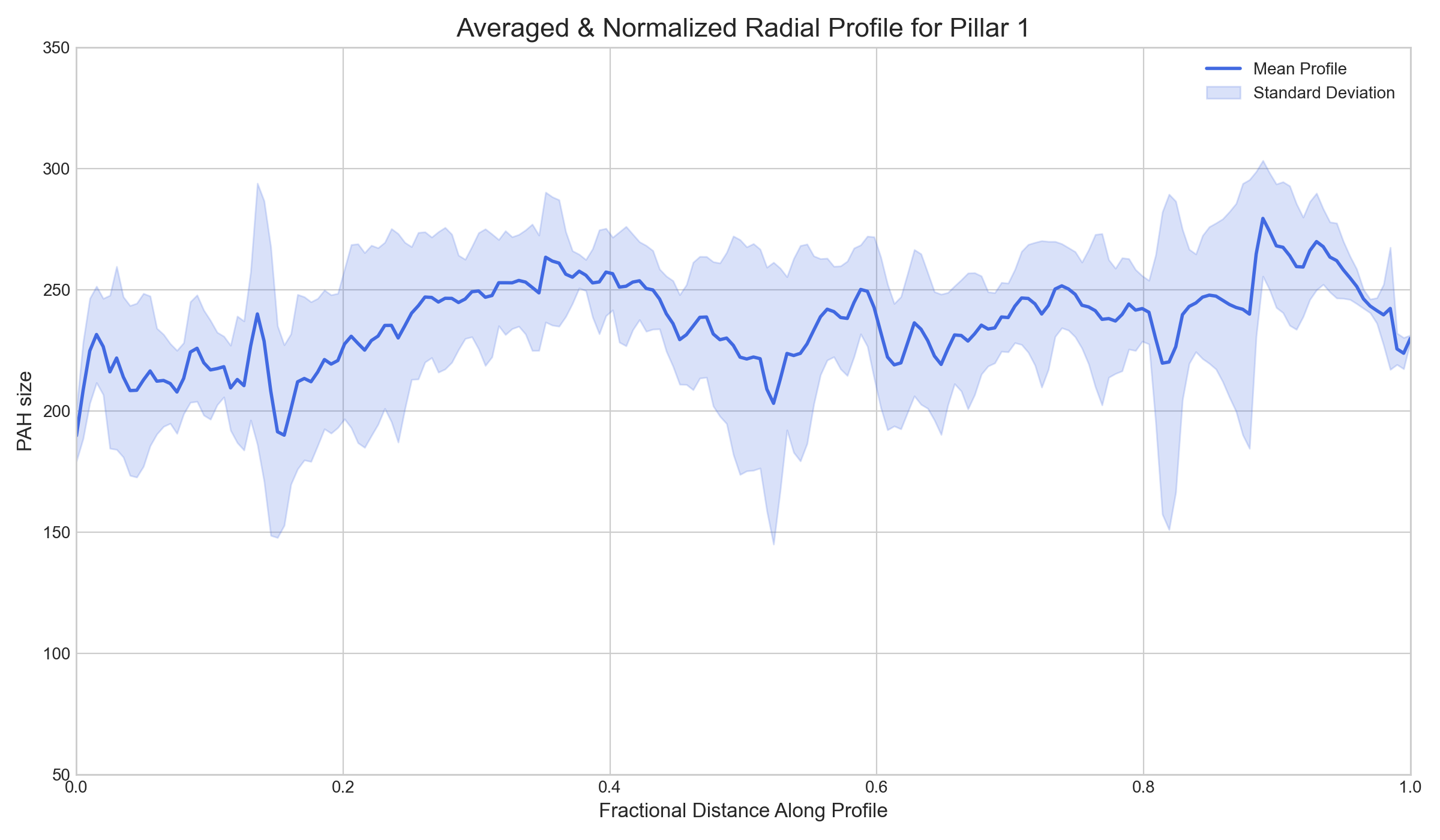}
    \caption{Averaged Radial Profile of Pillar 1a. Blue shaded region is the standard deviation of the 6 cuts.}
    \label{fig:p1}
\end{figure}

\begin{figure}
    \centering
    \includegraphics[width=1\linewidth]{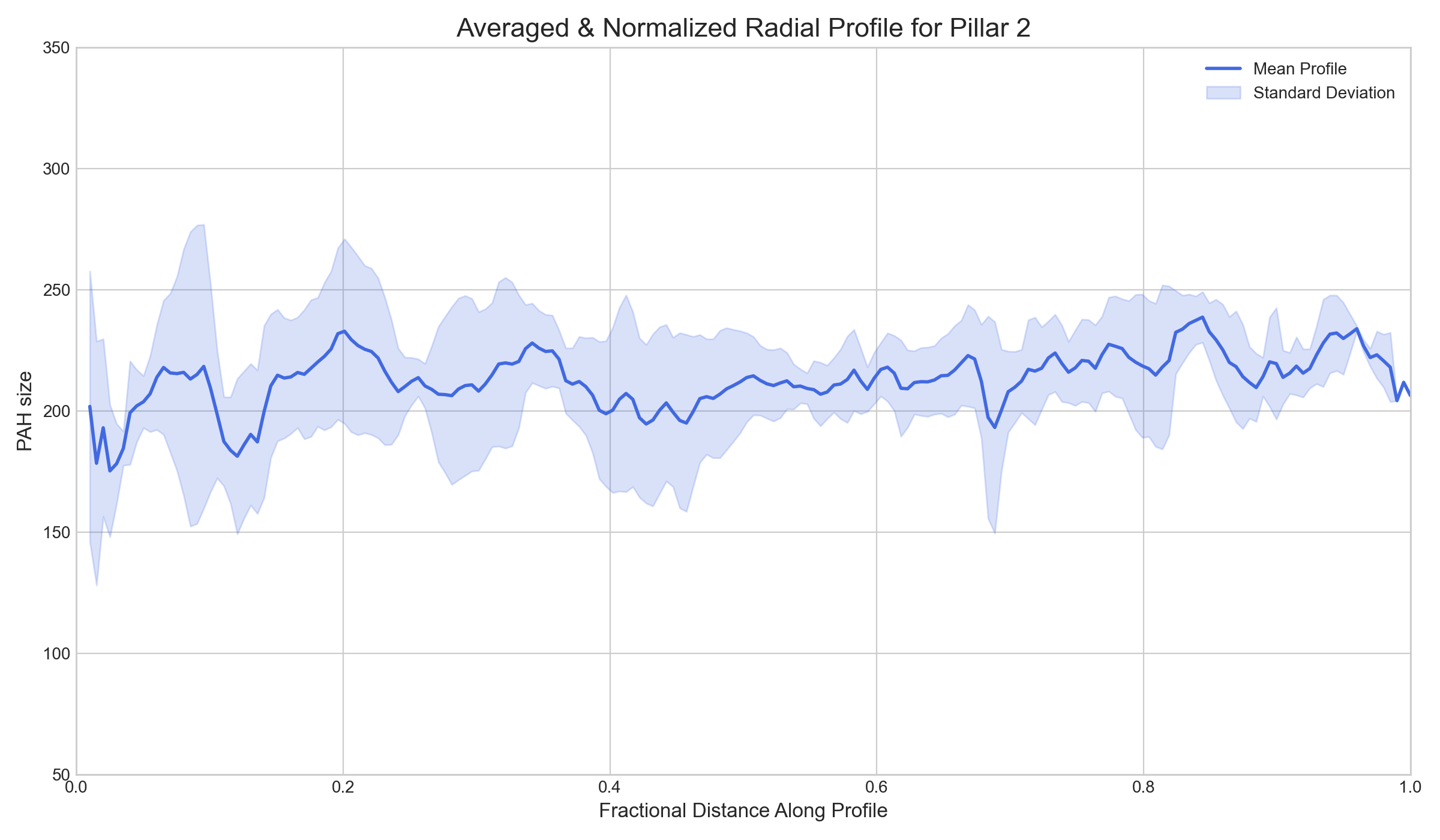}
    \caption{Averaged Radial Profile of Pillar 2. Blue shaded region is the standard deviation of the 6 cuts.}
    \label{fig:p2}
\end{figure}

To quantify the observed spatial variations, averaged and normalized radial profiles were extracted along multiple cuts for Pillar I (Figure \ref{fig:p1}) and Pillar II (Figure \ref{fig:p2}). These profiles track the evolution of the PAH size across the pillar, perpendicular to its axis. Both pillars exhibit a rather flat trend of PAH sizes as one moves across the pillar. The profiles also serve to elucidate physical differences between the pillars. Pillar II, has a profile at a lower PAH size ($\sim$200) compared to Pillar I ($\sim$250). 

\begin{table}[ht]
\centering
\caption{Mean PAH size with its standard deviation, and the mean PAH uncertainty.}
\label{tab:size_stats}
\begin{tabular}{lccccc}
\hline
\textbf{Region Name} & \textbf{PAH Size} & \textbf{Size Uncertainty} \\
\hline
Pillar 1a & 233.62$\pm$33.51  & 1.08 \\
Pillar 1b & 227.77$\pm$36.85  & 0.59\\
Pillar 2 & 212.73$\pm$29.45 & 1.22 \\
Molecular Cloud & 194.37$\pm$29.21 & 0.66 \\
\hline
\end{tabular}
\end{table}

\begin{table*}[ht]
\centering
\caption{Region-wise distribution of each ionization state}
\label{tab:ion_stats}
\begin{tabular}{lcccccc}
\hline
\textbf{Region Name} & \textbf{Mean ionization} &\textbf{Neutral} & \textbf{N75C25} & \textbf{N50C50} & \textbf{N25C75} & \textbf{Cationic} \\
 & (Ratio)& (\%) & (\%) & (\%) & (\%) & (\%) \\
\hline
Pillar 1a & 0.91$\pm$0.13& 65.21 & 33.12 & 1.55 & 0.11 & 0.00 \\
Pillar 1b & 0.88$\pm$0.14& 54.87 & 41.51 & 3.38 & 0.24 & 0.00 \\
Pillar 2 & 0.79$\pm$0.14& 25.12 & 68.31 & 6.03 & 0.5 & 0.04 \\
Molecular Cloud & 0.68$\pm$0.14& 0.20 & 75.00 & 21.08 & 3.36 & 0.36 \\
\hline
\end{tabular}
\end{table*}

Finally, a statistical analysis is conducted for different regions, with respect to their size and ionization. 4 regions were chosen: Pillar 1, Pillar 1a, Pillar 2, and Molecular cloud. The mean PAH size (Table \ref{tab:size_stats}) and ionization (Table \ref{tab:ion_stats}), with their standard deviation, were calculated. With this mean PAH size, and ionization uncertainty was also calculated. The small standard deviation indicates a uniform region. From Table \ref{tab:size_stats} it can be seen that Pillar 1a, which is the most irradiated region, has the highest size, while the molecular cloud has the smallest PAH size. Regions with lower radiation intensity have smaller PAH molecules.

From Table \ref{tab:ion_stats}, we can see that the regions with high radiation intensity, have more neutral molecules, and the mean ionization is closer to 1, indicating more neutral population. The majority population is either pure neutral or N75C25, with N75C25 dominating in regions that are comparatively more shielded (Pillar 2 and molecular cloud). The cationic population also increases in these regions. We can see a general trend that as one goes to regions with lower radiation intensity being more cationic, and regions that are more irradiated, showing a more neutral population.

\section{Discussion}\label{5}

This multi-wavelength analysis of the M16 pillars using the unparalleled sensitivity and resolution of JWST provides a detailed, spatially-resolved view of the lifecycle of Polycyclic Aromatic Hydrocarbons (PAHs) under the influence of intense stellar feedback. The results presented in the previous section allow us to construct a comprehensive physical narrative of how these interstellar molecules are processed, tracing their evolution from the shielded interiors of molecular clouds to the harsh, ionized environment of the HII region. The primary finding of this work is the strong spatial evidence for the size-dependent processing of the PAH population, a process fundamentally driven by the intensity and hardness of the local radiation field. Alongside this, the analysis also reveals the ionization properties of these molecules within the nebula.

\subsection{PAH Photo-processing}

One of the main observations from our various maps and plots is the effect of the FUV radiation from the stars of the NGC 6611 cluster on the PAH molecules. There is ample evidence of hard stellar radiation destroying PAH molecules (\cite{draine_infrared_2007}). This process has been confirmed through laboratory experiments (\cite{allain_photodestruction_1996}) and is observed on galactic scales as a decrease in the PAH-to-dust ratio in regions with intense radiation fields (\cite{galliano_variations_2008}). It has also been observed that larger PAHs are more photo stable than smaller ones (\cite{zhen_laboratory_2015}). 

This is consistent with the observations presented above. In the PAH size map (Figure \ref{fig:size}), regions having higher radiation intensity (Figure \ref{fig:f187}), harbor larger PAH molecules, and the more shielded the region is, the smaller the average size. The entire nebula is subject to intense radiation, the mean PAH size of $\sim$190 can be expected.

 The largely flat radial profiles indicate that no strong monotonic PAH-size gradient is resolved across the selected cuts. This is consistent with relatively homogeneous processing across the observed pillar surfaces, although projection effects, smoothing, and the choice of cut geometry may also dilute intrinsic gradients.

Region-wise analysis shown in Table \ref{tab:size_stats} shows clearly that the regions exposed to intense radiation, such as Pillar 1a, 1b have larger PAH molecules. As the intensity of the radiation drops by region, the average size of the molecules also drop, which again indicates photo processing as expected.

\subsection{Ionization}
The ionization map provides a spatially resolved view of how the PAH population responds to the local radiation field across the pillars. Distinct regions are dominated by different charge states, with large portions of the nebula exhibiting either neutral or mildly ionized (N75C25) PAHs. It can be seen from Figure \ref{fig:ion} that most of the regions in the nebula are either completely neutral or slightly ionized (N75C25 state). 

Ionization of PAH molecules is generally determined by the parameter $\gamma$ (\cite{tielens_photodissociation_1985}). It is a measure of the ratio of the rate at which FUV photons strike PAH molecules to the rate at which electrons recombine with these molecules. Smaller $\gamma$ represents a neutral population as there are enough electrons to recombine to maintain neutrality, and as $\gamma$ increases, the population becomes more ionic. 
\begin{equation}
    \gamma=\frac{G_0T^{1/2}}{n_e}
\end{equation}
Where $G_0$ is the FUV radiation in habing units, T is the gas temperature, and $n_e$ is the electron density. Using this the ionization of the regions of the nebula can be explained.

The entirety of Pillar 1a and the outer layer of Pillar 2 facing Pillar 1a (PDR) is almost purely neutral in ionization. This is shown quantitatively in Table \ref{tab:ion_stats}. This region, marked by a neutral population, has the highest radiation intensity in the entire nebula (Figure \ref{fig:f187}). The PDR on Pillar 2 has a more neutral population. This pattern is consistent with electron density playing a major role in setting the PAH charge balance, since high electron density enhances recombination and can offset the stronger radiation field. As shown by the MUSE data (\cite{leod_pillars_2015}), the electron density at the PDR surface and Pillar 1a is extremely high. This high $n_e$ dramatically increases the electron-ion recombination rate, which effectively overpowers the high photoionization rate from $G_0$. The net effect is a low ionization parameter $\gamma$, thus explaining the observed neutral state of the PAH population precisely where the stellar irradiation is strongest. In summary, the high electron density and low temperature in the region facilitates recombination, maintaining a neutral population. 

The molecular cloud has a negligible neutral population and is more cationic compared to the other regions (Table \ref{tab:ion_stats}).  I can compare the $\gamma$ values between the Pillar 1 and molecular cloud, which is expected to be lesser than 1. $G_0$ (value taken from F187N image) ratio can be got by taking the mean value of the F187N image in the respective regions and converting to habing units, and it is approximately;
\begin{equation}
    \frac{G_0(Pillar)}{G_0(Mol cloud)}\sim2
\end{equation}

From \cite{flagey_tracing_2011}, the temperature of the molecular cloud is 32K and that of the nebula/pillars is 40K, hence their ratio is approximately 1. Putting these into the equation;
\begin{equation}
    \frac{\gamma(Pillar)}{\gamma(Mol Cloud)}=\frac{n_e(Mol Cloud)}{n_e(Pillar)}*2
\end{equation}

From the ionization map, we see that the molecular cloud is cationic compared to Pillar 1. Hence; 
\begin{equation}
    \frac{n_e(Pillar)}{n_e(Mol Cloud)}>2
    \label{constraint}
\end{equation}
From the electron‐density map of McLeod et al. (their Fig. 8a; \cite{leod_pillars_2015}), the $n_e$ values reach $>$2000 $cm^{-3}$ of Pillar 1a at the tip, while the pillar body typically lies at a few hundred $cm^{-3}$. Adopting a characteristic, area-averaged value for Pillar 1a of $n_{e, P1} \sim 1000 - 1500 cm^{-3}$, I get the approximate upper limit of the electron density in the molecular cloud to be $n_{e, cloud} \sim 300-700 cm^{-3}$
, i.e., roughly a factor of two lower. These numbers are order-of-magnitude, literature-anchored estimates based on the published [S II] diagnostic map and are not precise regional means. Equation \ref{constraint} gives a rough constraint on the electron density. This is expected as in the molecular cloud, the intensity of radiation is much lower, so
ionization mainly happens through cosmic rays and a
few scattered rays (\cite{tielens2005}), hence the ionization is weak and
much slower, leading to a lower electron density.

One interesting feature to note is the region on top of Pillar 1a. Given that the region is present directly under the star cluster, it should be the region experiencing the highest intensity radiation. According to the above logic, we would expect it to be neutral. It can be seen from Figure \ref{fig:ion} that the region on top of Pillar 1a has an ionization state of N75C25, harbouring a slightly cationic population. From Figure \ref{fig:size}, it is evident that this region has a lower PAH size compared to its surroundings. This suggests that this region is not strongly irradiated as the other parts of Pillar 1a. This is due to the density of gas in this region. Near-IR images (\cite{Sugitani_2002}) reveal that, instead of being dense
continuous structures of gas and dust, Pillar 1a and 2 are made of relatively low-density material, capped by much denser cores.One possible explanation is that the dense cap seen in earlier near-IR work provides stronger shielding, reducing the net photo-processing efficiency in this region. However, without direct local constraints on density and temperature, this remains tentative.

\subsection{Relation between Ionization and sizes}

It is worth to explore the relation between ionization and sizes. A Spearman correlation coefficient of 0.58 was found, indicating a moderately strong correlation. Figure \ref{fig:sizedis} shows a normalized histogram of PAH sizes by ionization state. It can be seen that the neutral population has the highest PAH sizes, with a mean size of 240, followed by N75C25 with a mean size of 214.7. The smallest PAH size is of the cationic population, with a mean size of just 60 carbon atoms. There is a clear trend of ionized PAH molecules being smaller in size than neutral population. 

The cationic population is present in regions with lower radiation intensity, hence their size is smaller, while the neutral molecules are present in regions with higher intensity, leading to the destruction of the smaller molecules while the larger molecules survive. As one increases the cationic fraction, the size of the molecules decreases. Although the ionization potential (IP) of PAHs decreases with increasing molecular (\cite{montillaud_evolution_2013}) size, meaning larger PAHs require less energy to ionize, the observed trend is that smaller PAHs dominate the ionized population in high-$\gamma$ regions. This apparent contradiction can be explained by considering the interplay between ionization, fragmentation, and survival bias. Smaller PAHs, despite having higher IPs, possess fewer vibrational modes and are therefore less able to dissipate absorbed UV energy non-destructively \cite{allain_photodestruction_1996}. Once exposed to a radiation field, depending on the intensity they are more likely to undergo ionization or, in cases of higher intensity destroyed entirely through photo-fragmentation (\cite{zhen_laboratory_2015}). Large PAHs, on the other hand, are more stable as they can dissipate the energy through vibrational modes and tend to survive as neutral molecules. Given their location in high electron density regions, even if they do ionize, they quickly recombine with surrounding electrons to maintain neutrality. As a result, the neutral PAH population skews toward larger sizes, while the cationic population becomes dominated by the surviving small PAHs. This leads to the observed anticorrelation between PAH size and ionization fraction, even though large PAHs are energetically easier to ionize.

\section{Comparison}\label{6}
The observed trend of increasing average PAH size in the irradiated PDRs of M16 is consistent with findings in other high-radiation environments. For instance, \cite{knight_tracing_2021} investigated the reflection nebulae NGC 7023 and NGC 2023 and similarly found that the average PAH size, as traced by the 11.2/3.3 $\mu$m ratio, increases with proximity to the illuminating stars. This provides strong corroborating evidence that the photo-processing of PAHs, where smaller molecules are preferentially destroyed, is a common and dominant mechanism in PDRs.

In contrast to these high-energy environments, recent JWST observations of the Ring Nebula (NGC 6720) by \cite{clark_jwst_2025} found a PAH population that is largely neutral and dominated by small molecules (approximately 35 carbon atoms). The Ring Nebula represents a different evolutionary stage and radiation environment, and it serves as a useful baseline, highlighting that the large, processed PAHs seen in M16 are a direct consequence of the extreme feedback from the massive stars of the NGC 6611 cluster.

The mean size of PAH in M16 is $\approx$193, which is much higher than the value of 35 in Ring Nebula. This difference arises because the Ring Nebula is a planetary nebula (\cite{clark_jwst_2025}, where PAHs are likely recently formed and dominated by small species, whereas M16 is a massive star-forming region. The intense and pervasive UV radiation from the NGC 6611 cluster preferentially destroys the smallest PAHs, leaving behind a population skewed toward larger molecules.

\section{Limitations and caveats}

This study is based on filter-derived PAH diagnostics rather than spatially resolved mid-infrared spectroscopy, and this is the principal limitation of the analysis. The PAH size and charge-state framework adopted here follows the methodology of M20, but that work was developed and calibrated using spectroscopic measurements, which provide more direct and accurate separation of PAH bands from the underlying continuum and from neighbouring spectral features. In M16, comparable spatially resolved spectroscopy is not presently available at the required coverage and resolution, and we therefore rely on JWST imaging to construct approximate PAH feature maps. The resulting size and ionization estimates should therefore be interpreted as imaging-based proxies rather than unique measurements of intrinsic molecular properties.

A second limitation arises from the continuum subtraction itself. The continuum at each PAH band was estimated using interpolation between shorter and longer wavelength anchor filters, but in several cases, the available anchors are spectrally distant from the PAH feature of interest. This can introduce uncertainty if the true continuum shape deviates from the assumed local power-law form. As already noted in the methods, this may lead to incomplete subtraction or systematic residuals, especially in regions where the dust continuum varies rapidly across the nebula.

Related to this, the continuum interpolation assumes a single power-law behavior in log-log space between the two anchor filters. While this is a practical approximation, the true mid-infrared continuum in a complex H II region / PDR environment can include curvature \cite{Compi_gne_2008}, multiple dust components, and line contamination that are not captured by such a simple form. Any departure from the assumed continuum shape will propagate directly into the continuum-subtracted PAH maps and hence into the derived ratio-based diagnostics.

Finally, the nebula is a projected three-dimensional structure, and line-of-sight mixing may blur the connection between observed surface brightness ratios and the true local physical conditions. Regions that appear spatially distinct in projection may contain overlapping material with different irradiation histories, densities, and PAH populations. The derived maps should therefore be interpreted as projected diagnostics of the dominant emitting material along each line of sight, rather than as direct measurements of a single homogeneous layer.

Despite these limitations, the main conclusions of this work remain meaningful because they rely primarily on relative spatial trends rather than on the absolute calibration of any single pixel. The derived PAH maps show spatially coherent and regionally ordered behavior. The resu;lts can be interpreted as spectroscopic-grade absolute measurements of PAH properties, but as robust imaging-based constraints on how PAH populations vary across the M16 environment.

\section{Conclusion}
I have presented a comprehensive spatial analysis of the polycyclic aromatic hydrocarbon (PAH) population across the Pillars of Creation in M16, using high-resolution imaging from JWST’s MIRI and NIRCam instruments. Our results provide insight into how PAHs respond to intense stellar feedback and evolve across complex photodissociation regions.

I constructed detailed maps of PAH size, ionization state, and abundance, and examined their relationship to the local radiation field and gas content. The PAH size map, derived from the F1130W/F335M ratio, reveals that larger PAHs are concentrated along the pillar surfaces and irradiated edges, consistent with their increased stability in UV-rich environments. Conversely, smaller PAHs dominate the interior and shielded regions, pointing to either fragmentation or inhibited growth in low-irradiation zones. The ionization map, based on the grid and modeled using 10eV charge-state diagnostics, shows the dominance of neutral population in the pillars and the population becoming cationic in the shielded regions. This can be explained by looking at the balance between ionization and recombination. Using this, we were able to predict a rough upper limit for the electron density in the molecular cloud to be about half of that in the Pillars. 

The RGB composite image combining F187N (ionized gas), F470N (molecular gas), and F770W (PAHs) illustrates the layered morphology of the pillars and highlights key PDR interfaces. The normalized radial profiles along 6 cuts across the two pillars show a quite flat profile, indicating a uniform distribution of PAH and radiation across the pillar. The normalized histogram showing the distribution of PAH size by ionization shows the anti-correlation between the cationic fraction and size. 

Together, these results demonstrate that the PAH population in M16 is highly sensitive to local radiation and density structure, with measurable gradients in size, abundance, and ionization across sub-parsec scales. This study also underscores the power and limitations of filter-based diagnostics in extreme environments, and sets the stage for future spectroscopic follow-up and observations with more filters. The methodology developed here can be extended to other star-forming regions to investigate how PAHs evolve under different feedback conditions, providing critical constraints on their lifecycle and role in the ISM.

\section*{Data Availability}
The JWST data used in this work are available from the MAST archive with programme ID 2739. 
Derived data products and analysis scripts are available from the corresponding author upon reasonable request.

\bibliographystyle{mnras}
\bibliography{citations}

\end{document}